\documentclass[copyright,creativecommons]{eptcs}
\providecommand{\event}{AMCA-POP 2010} 

\providecommand{\firstpage}{1}

\usepackage{algorithmic}
\usepackage{algorithm}
\usepackage{breakurl}             
\usepackage{amsmath}
\usepackage{amsthm}
\usepackage{amssymb}
\usepackage{amsfonts}
\usepackage{mathrsfs}
\usepackage{multirow}
\usepackage{subfigure}
\usepackage{stmaryrd}
\usepackage{listings}
\usepackage{booktabs}
\usepackage[all]{xy}

\usepackage{enumerate}
\usepackage{subfigure}
\catcode`\"=\active
{\gdef"{\begingroup\ttfamily\let"=\endgroup}}

\newif\ifdraft
   \drafttrue

\newif\ifpdf
        \pdffalse

\ifpdf
        \usepackage[pdftex]{graphicx}
\else
        \usepackage[dvips]{graphicx}
        \usepackage[dvips]{color}
\ifdraft
        \usepackage[dvips,light,none]{draftcopy}
\draftcopyName{$Revision: 281 $}{120}
\fi
\fi

\usepackage{colortbl}
\usepackage{pstricks}
\usepackage{pst-node}

\def\cA{{\cal A}}

\def\cG{{\cal G}}

\def\relset{\mathbb{Z}}

\def\Alleles{\cA}

\def\Maps{\mathscr{M}}

\def\Genotypes{\cG}
\def\mate{\mathsf{mate}}

\def\father{f}
\def\mother{m}
\DeclareMathOperator{\dom}{dom}
\newcommand{\Ped}[3]{\langle {#1}, {#2}, {#3} \rangle}

\newcommand{\modif}[2]{[#1/#2]}
\DeclareMathOperator{\LG}{\mathsf{LG}}
\DeclareMathOperator{\OCW}{\mathsf{OCW}}
\DeclareMathOperator{\OCWR}{\mathsf{OCWR}}
\DeclareMathOperator{\Split}{\mathsf{split}}

\DeclareMathOperator{\Filter}{\mathsf{filter}}

\newcommand{\Clone}[1]{\overline{#1}}

\def\powerset{\wp}

\renewcommand\le{\leqslant}

\def\partial{\multimap\to}

\newcommand\TheTool{\ensuremath{\mathsf{Celer}}}

\theoremstyle{definition}
\newtheorem{definition}{Definition}
\theoremstyle{theorem}

\newtheorem*{proposition-no-number}{Proposition}

\newtheorem*{corollary-no-number}{Corollary}
\theoremstyle{remark}

\newtheoremstyle{example}{\topsep}{\topsep}%
{}
{}
{\bfseries}
{.}
{5pt}
{\thmname{#1}\thmnumber{ #2}\thmnote{ #3}}

\theoremstyle{example}
\newtheorem{example}[definition]{Example}

\begin{document}
\DeclareGraphicsExtensions{.jpg, .png, .pdf, .eps}

\title{\TheTool: an Efficient Program for Genotype Elimination}

\author{Nicoletta~De Francesco \and Giuseppe~Lettieri \and Luca~Martini
  \institute{Dipartimento di Ingegneria dell'Informazione: Informatica, Elettronica, Telecomunicazioni\\
    Universit\`a di Pisa\\
    Largo Lucio Lazzarino, 2\\
    56122 Pisa - Italy
  }
  \email{\{n.defrancesco,g.lettieri,luca.martini\}@iet.unipi.it}
}

\def\authorrunning{N.~De Francesco et al.}
\def\titlerunning{\TheTool: an efficient program for genotype elimination}
\maketitle

\begin{abstract}
  This paper presents an efficient program for checking Mendelian consistency in
  a pedigree. Since pedigrees may contain incomplete and/or erroneous
  information, geneticists need to pre-process them before performing linkage
  analysis. Removing superfluous genotypes that do not respect the Mendelian
  inheritance laws can speed up the linkage analysis. We have described in a
  formal way the Mendelian consistency problem and algorithms known in
  literature. The formalization helped to polish the algorithms and to find
  efficient data structures. The performance of the tool has been tested on a
  wide range of benchmarks. The results are promising if compared to other
  programs that treat Mendelian consistency.
\end{abstract}


{\bf keywords}: abstract interpretation 
\section{Introduction}
\label{sec:intro}
Geneticists employ the so-called \emph{linkage analysis} to relate genotypic
information with their corresponding phenotypic information.  Genotypes are
organized in data structures called \emph{pedigrees}, that besides genetic data,
record which individuals mate and their offspring. Since pedigrees may contain
incomplete and/or erroneous information, geneticists need to pre-process them
before performing linkage analysis.  Moreover, in many cases, we cannot know any
genetic information for some individuals (for instance because they refuse to or
cannot be analyzed) and we would like to know which are their possible
genotypes. Therefore, we would like to pre-process the pedigree by removing some
candidate genotypes, in such a way that the remaining genotypes respect the
classical Mendelian laws. When the pedigree is composed by thousands of
individuals, this consistency checking need to be automated. The first notable
contribution in the pedigree consistency check is the algorithm proposed by
Lange and Goradia in 1987~\cite{LangeGoradia1987}. The algorithm takes as input
a pedigree with a list of genotypes associated to every individual, and perform
genotypes elimination by removing from the lists the genotypes that lead to
Mendelian inconsistencies. The algorithm performs a fixpoint iteration by
processing one nuclear family at a time. This algorithm is optimal (in the
sense that it removes all the genotypes that lead to Mendelian
inconsistencies, and only them) when the pedigree has no loops. An example of
loop in a pedigree is 
when two individuals that mate have an ancestor in common. An algorithm that is
optimal even in the presence of loops has been proposed by O'Connell and Weeks
in 1999~\cite{OConnellWeeks1999}. In brief, the algorithm selects the loop
breakers (that is the individuals that, if duplicated, remove the loop) and
perform the Lange Goradia algorithm for every combination of the genotypes of
the loop breakers. Unfortunately, it has been proven~\cite{Aceto2004} that the
consistency check on pedigrees with marker data containing at least three
alleles is a NP-hard problem.

The remainder of the paper is organized as follows. Firstly, we formalize the
problem of genotype elimination (Section~\ref{sec:problem}) and the algorithms of
Lange-Goradia (Section~\ref{sec:lg}) and O'Connell and Weeks
(\ref{sec:ow}). Then, we describe the implementation of {$\TheTool$}
(Section~\ref{sec:imple}). Section~\ref{sec:performances-thetool} describes the
performances of {$\TheTool$} on a large set of benchmarks. Then, we compare our
program with other existing software (Section
\ref{sec:comp-with-other}). Finally, we conclude and suggest some directions for
future works.

\section{Mendelian consistency algorithms}
\label{sec:problem}

\begin{figure}
  \centering
  \includegraphics[width=0.3\linewidth]{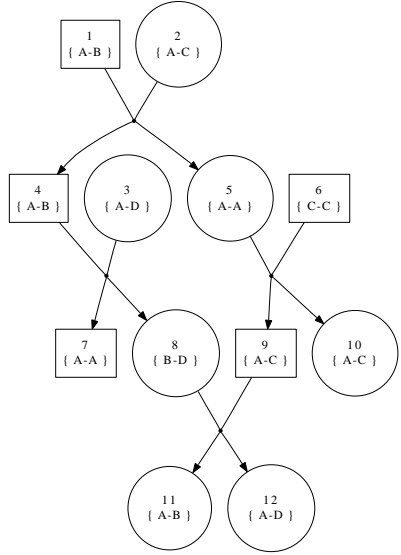}
  \caption{An example of a pedigree}
  \label{fig:OUT}
\end{figure}

A pedigree contains parental and genetic information about a set of individuals.
Pedigrees are usually represented in a graphical way by drawing a circle for
every female individual and a box for every male individual. Inside the circle
(or the box) there can be some data regarding the individual (for instance
genetic information, or affection status). Parental relations are represented by
lines that connect to a node (the so called marriage node). Arrows depart from
the marriage nodes to the children of the couple. In Figure~\ref{fig:OUT} we
report the graph of a pedigree composed by 11 individuals. For each individual,
we report his/her identification number (id from now on) and his/her possible
genotypes.

We collect the
parental structure in a triple $\Ped I f m$ where $I$ is the set of individuals
and $f$ and $m$ are two partial functions from $I$ into $I$ mapping a subset
$\dom f = \dom m \subset I$ of individuals to their father and mother,
respectively.  The individuals that do not have parents in the pedigree are
called \emph{founders}. For the pedigree of Fig.~\ref{fig:OUT}, the founders are
the individuals with id in $\{1,2,3,6\}$.

We suppose that we are looking at a single locus. The possible alleles in the
locus are in the set $\Alleles$, ranged by uppercase case letters $A, B, C,
\ldots$. Let $\Genotypes$ be the set of unordered pairs of elements in
$\Alleles$. Since we consider the genotypes $(A,B)$ and $(B,A)$ as equivalent,
the genotype of each individual will be an element of the set $\Genotypes$. A
fully specified genetic map of a pedigree $\Ped I f m$ is an element $h$ of
$I\to \Genotypes$.  We say that a fully specified map (fsmap from now on) is
Mendelian if the genotypes of every non-founder individual is such that one of
its allele is derived from the mother and the other from the father.  It is
often useful to check for Mendelian consistency in a subset of the individuals
in the pedigree. Since the Mendelian conditions involve an individual and both
his parents, it makes sense to consider those subsets that contain either both
or none of the parents of each individual in the subset. Given a pedigree $\Ped
I m f$ we say that $S\subseteq I$ is a {\em regular} subset of $I$ if, for each
$i\in \dom f \cap S$, we have that $f(i) \in S \iff m(i) \in S$. Intersections
and unions of regular subsets are again regular subsets. For instance, in the
pedigree of Fig.~\ref{fig:OUT}, the set $\{3,4,7,8,9,11,12\}$ is an an example
of a regular subset of the individuals.

We can also define a function
$\mate\colon\Genotypes\times\Genotypes\to\powerset(\Genotypes)$ that, given two
genotypes, returns the set of Mendelian genotypes that can be generated by
selecting one allele from each one. We have (remember that we use unordered
pairs):
\begin{displaymath}
  \mate((A,B)(C,D)) = \{(A,C),(A,D), (B,C), (B,D)\}
\end{displaymath}
With the help of function $\mate$, we can now express more precisely when a
fsmap is Mendelian on a regular subset of individuals:
\begin{definition}[Mendelian consistency]
  Let $P=\Ped I f m$ be a pedigree and let $S$ be a regular subset of $I$.
  The fully specified map $h$ is {\em Mendelian on $S$} if and only if for every individual
  $i \in S$ such that $\father(i)\in S$ and $\mother(i)\in S$, we have $h(i)\in
  \mate(h(\father(i)),h(\mother(i)))$.
\end{definition}
We say that an fsmap $h$ on a pedigree $P=\Ped I f m$ is {\em Mendelian} if it
is Mendelian on $I$. The reader can verify that the fsmap in Fig.~\ref{fig:OUT} is Mendelian.

Since in general we do not know precisely the genotype of each individuals, only
partially specified maps will be available. A partially specified map $H$ (psmap from now on)
records for every individual of the pedigree the genotypes it may have according
to our information (e.g. because we have collected some genetic data or we have
observed the phenotype). A psmap $H$ is an element of the set
$I\to\powerset(\Genotypes)$. We can introduce a partial
order relation $\sqsubseteq$ on set $\Maps$. We say that map $H_1$
is more precise than or equal to map $H_2$, and we write $H_1 \sqsubseteq H_2 $,
if and only if, for every individual $i\in I$, $H_1(i) \subseteq H_2(i)$.
With an abuse of notation we identify any fully specified map $h$
with the partially specified map that maps $\{h(i)\}$ to every individual
$i\in I$.
Thus we write $h\sqsubseteq H$ to mean that, for every individual $i$, $h(i)\in H(i)$. 
All psmaps such that $H(i) = \varnothing$ for any $i \in I$ describe an inconsistent situation
where no possible assignment of genotypes is compatible with the available information.
We identify all these psmaps and denote them by $\bot$,
the psmap that maps $\varnothing$ to all individuals in $I$.
We denote by $\Maps = (I\to\wp(\Genotypes))/\bot$ the set obtained by this identification.
The set $\Maps$ is a complete lattice, with least upper bound $\bigsqcup$ given
by pointwise union. 
The greatest lower bound $\bigsqcap$ is obtained in two steps: first,
the pointwise intersection is computed; then, if any individual is mapped to $\varnothing$
in the previous step, the result is taken to be $\bot$.

In psmaps we are interested in those genotypes, taken from the sets of 
each individual, that can be used to build a Mendelian fsmap.

\begin{definition}[Consistent genotype]
  Let $P=\Ped I f m$ be a pedigree and let $S$ be a regular subset of $I$.
Given a psmap $H$ and an individual $i\in I$, we say that
genotype $g\in H(i)$ is {\em consistent on $S$} if{}f there exists
an fsmap $h \sqsubseteq H$ with $h(i) = g$ such that $h$ is Mendelian on $S$.

A psmap $H$ is consistent on $S$ if all $g\in H(i)$, for all $i\in I$, are consistent on $S$.
\end{definition}

A pedigree consistency algorithm can be seen as a function that takes a psmap
and returns another psmap where some inconsistent genotypes have been removed.
More precisely, we define function $\Filter_S\colon \Maps \to \Maps$ such that
$\Filter_S(H) = H' \sqsubseteq H$ and $H'$ is consistent on $S$.

We say that a psmap $H$ on a pedigree $\Ped I f m$ is {\em fixed} on a set $S\subseteq I$
if $H(i)$ is a singleton set for all $i \in S$.
\begin{example}
  \label{ex:filter-on-nuclear-family}
Let $i\in I$ be a non-founder 
in the pedigree $\Ped I f m$ and assume the psmap $H$ 
is fixed on $\{f(i), m(i)\}$. Thus $H(f(i))=\{g_f\}$ and $H(m(i))=\{g_m\}$.
Let us compute $H'=\Filter_{\{f(i), m(i), i\}}(H)$. 
Consider $G=H(i)\cap \mate(g_f, g_m)$. If $G\neq\varnothing$ then $H'=H[G/i]$,
otherwise $H' = \bot$.
\end{example}

Let $S$ and $T$ be two regular subsets of $I$. 
We may want to obtain $\Filter_{S\cup T}(H)$
from $\Filter_S(H)$ and $\Filter_T(H)$, which may be simpler to compute.
A candidate composition is $\Filter_S(H)\sqcap\Filter_T(H)$, since this operation
keeps the genotypes which are consistent on both $S$ and $T$.
However, in general, we only have
$\Filter_{S\cup T}(H)\sqsubseteq\Filter_{S}(H)\sqcap\Filter_{T}(H)$,
and the relation may be strict.
Nonetheless, it can be easily seen that the
equality holds whenever 
$H$ is fixed on $S\cap T$.

A useful function in the definition
of consistency check algorithms is function $\Split_S\colon \Maps\to\wp(\Maps)$.
Given any $S\subseteq I$, $\Split_S(H)$ is the set of all psmaps $F\sqsubseteq H$
such that $F$ is equal to $H$
on $I\setminus S$ and is fixed on $S$.
Thus, if $S=\{x_1, \ldots, x_n\}$, then
for each $(g_1, \ldots, g_n)\in H(x_1)\times\cdots\times H(x_n)$
we have a psmap $F\in\Split_S(H)$ such that $F(x_i) = g_i$ for all $1\le i \le n$
and $F(x) = H(x)$ for all $x\not\in S$.
If $P=\Ped I f m$ is a pedigree and $H$ is a psmap on it,
we have the following relation for all $T, S \subseteq I$ (where $S$ is regular)
\begin{equation}
  \label{eq:split-filter-sup}
  \bigsqcup_{F\in\Split_T(H)} \Filter_S(F) = \Filter_S(H).
\end{equation}

\begin{figure}
  \centering
  \subfigure[]{\includegraphics[width=.32\linewidth]{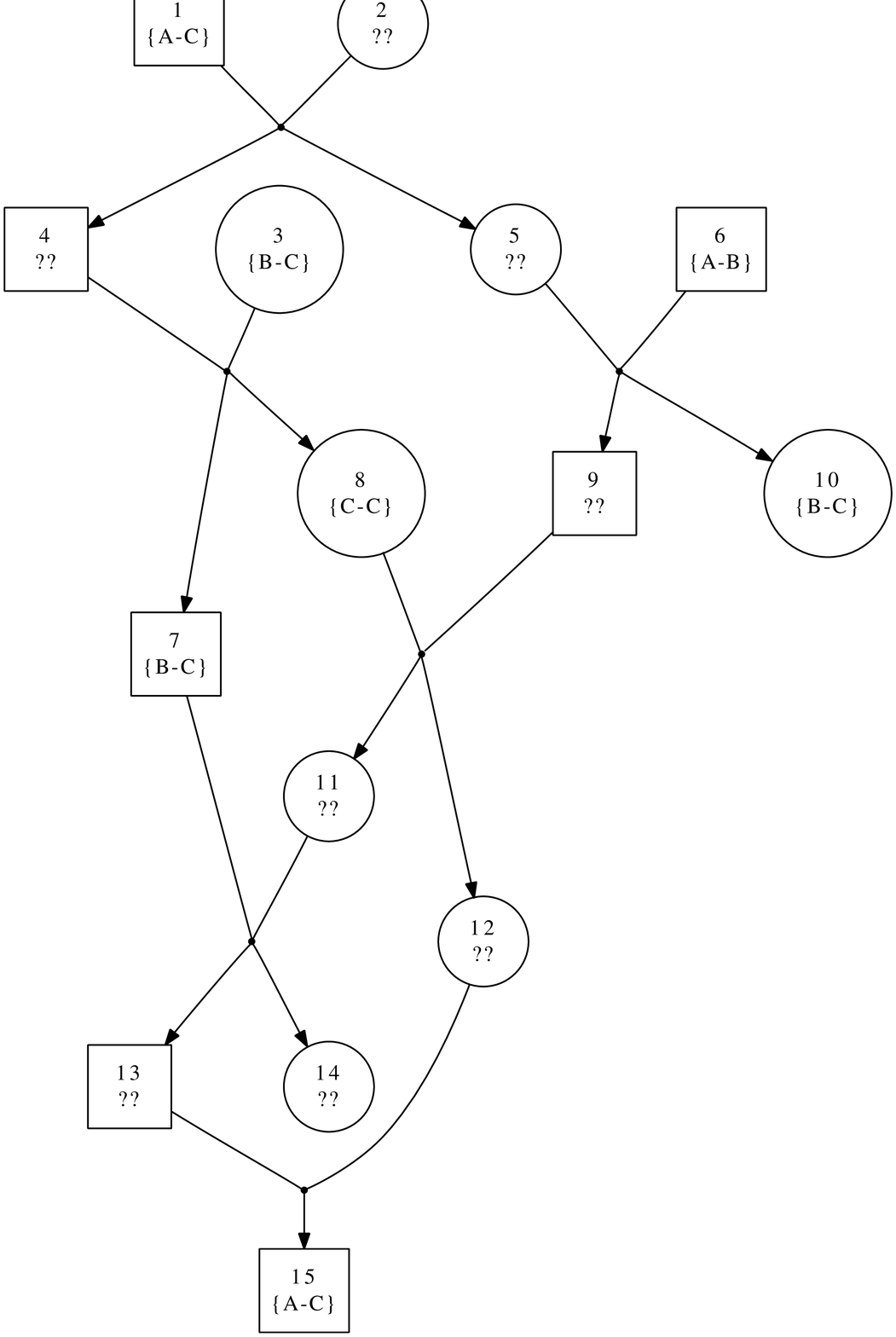}\label{fig:subfigex2}}
  \subfigure[]{\includegraphics[width=.32\linewidth]{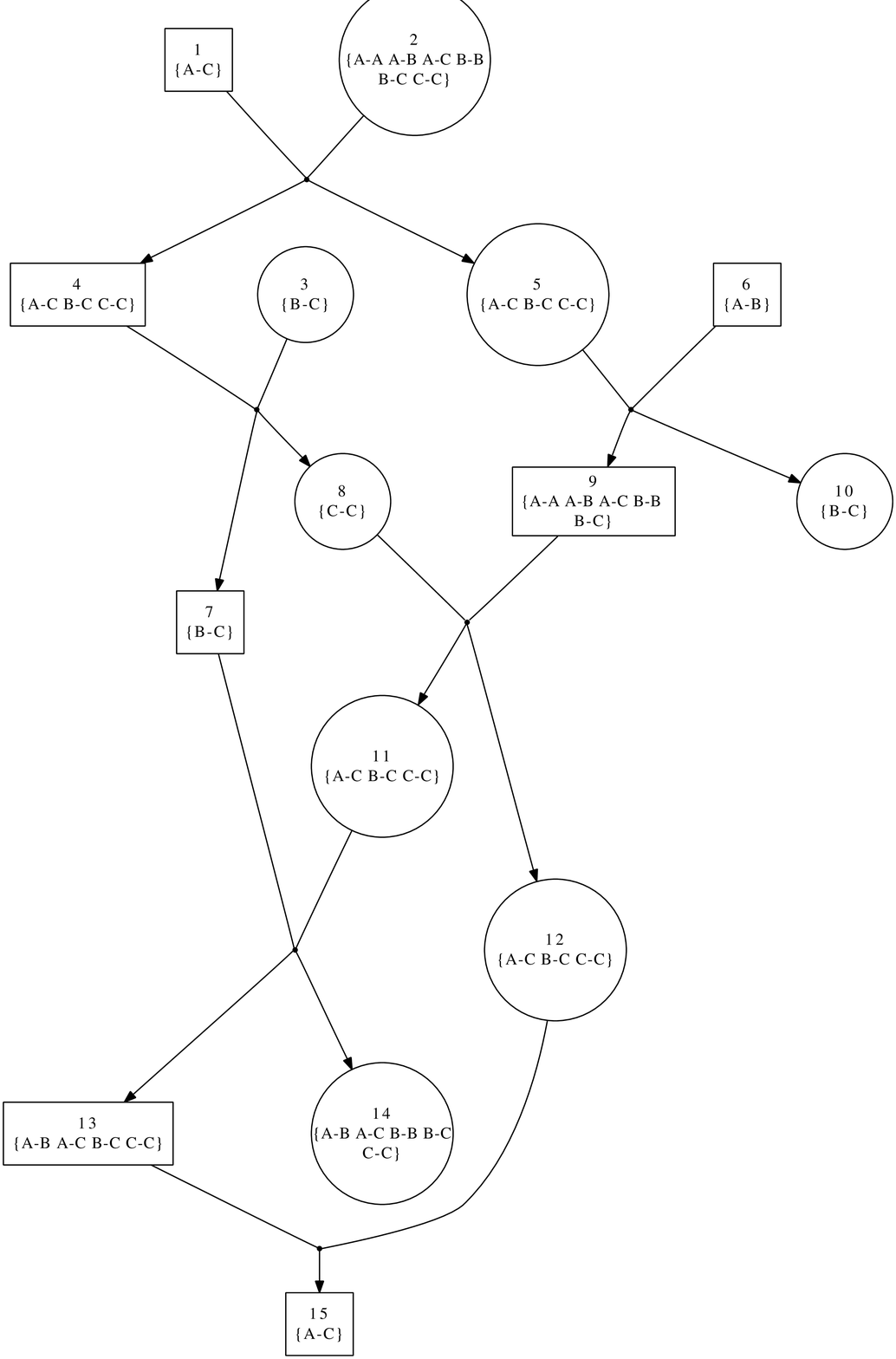}\label{fig:subfigex2lg}}
  \subfigure[]{\includegraphics[width=.32\linewidth]{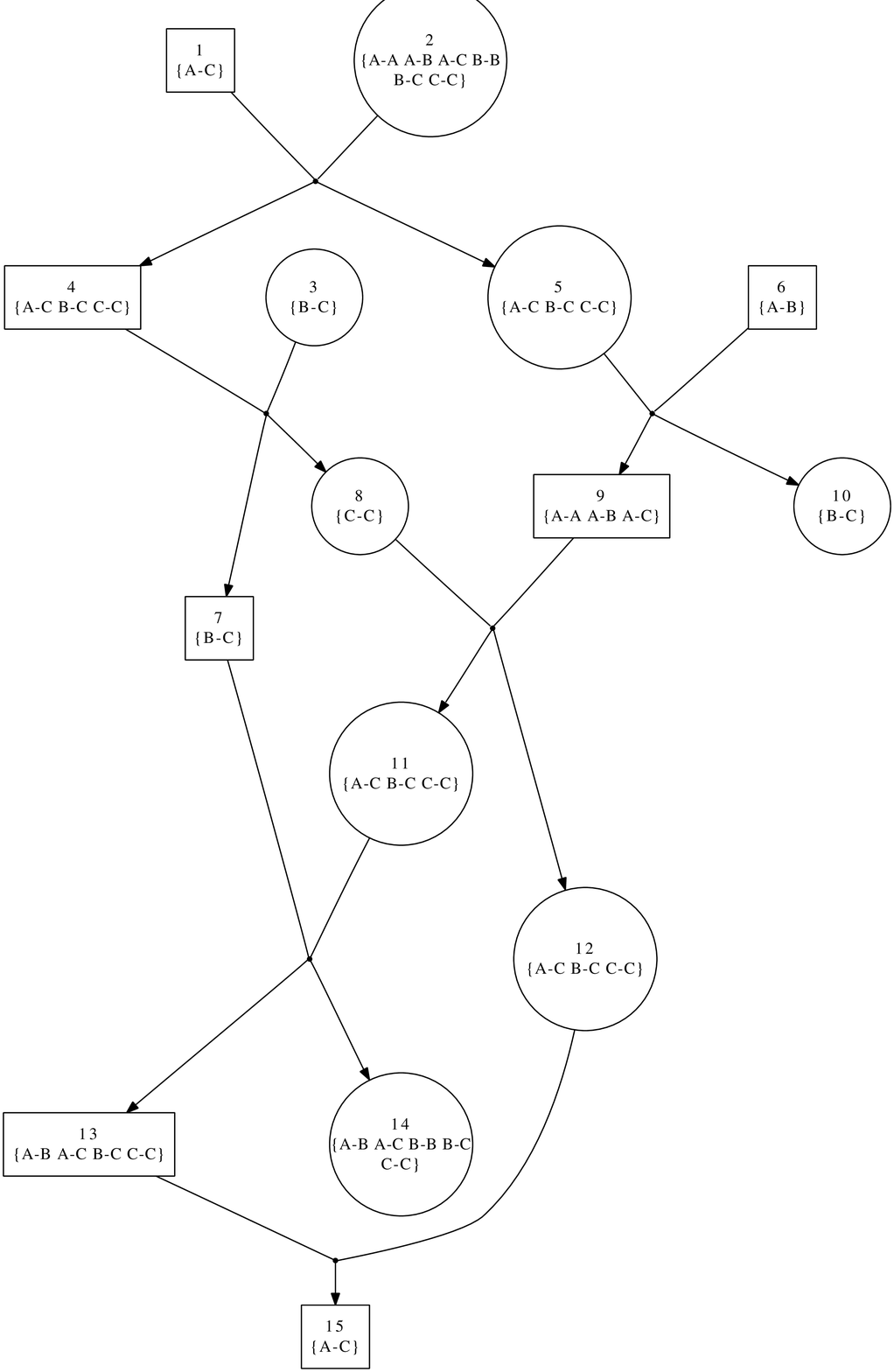}\label{fig:subfigex2ow}}
  \caption{An example of the applications of the genotype elimination
    algorithms: the initial pedigree \ref{fig:subfigex2}, after the application
    of Lange-Goradia algorithm \ref{fig:subfigex2lg}, and after the application
    of O'Connell and Weeks \ref{fig:subfigex2ow}. In the initial pedigree, we
    have marked with ``??'' the untyped individuals.}
  \label{fig:example2}
\end{figure}



\subsection{The Lange-Goradia algorithm}
\label{sec:lg}

The idea of the Lange-Goradia algorithm is to remove all the genotypes of an individual $i$
that are inconsistent on any nuclear family to which $i$ belongs.
This is accomplished by looking at one nuclear family at a time.
Let $H$ be a psmap for a pedigree $\Ped I f m$.
If $S=\{x,y,k_1,\ldots,k_n\}\subseteq I$ is a nuclear
family where $x$ and $y$ are the parents and $k_1, \ldots, k_n$
are the children, then each pair $(g_x, g_y)$
of genotypes in $H(x)\times H(y)$
is examined in turn, checking that $\mate(g_x, g_y) \cap H(k_i) \neq \varnothing$
for all the children $k_i$ with $i = 1, \ldots,n$.
If this is the case, then $g_x$, $g_y$ and all genotypes in $\mate(g_x, g_y)\cap H(k_i)$
for each children $k_i$ are consistent on $S$. 
All genotypes that are not found to be consistent after all pairs
of genotypes in $H(x)\times H(y)$
have been examined are certainly inconsistent on $S$ and, thus, also inconsistent, so they
can be safely removed.
More formally, we can say that the algorithm computes
$\bigsqcup_{F\in\Split_{\{x,y\}}(H)}\Filter_S(F)$ (note that a nuclear family
is a regular subset of $I$), which is equal to $\Filter_S(H)$ according to \eqref{eq:split-filter-sup}. For each $F\in\Split_{\{x,y\}}(H)$, $\Filter_S(F)$ is computed
as $\bigsqcap_{i=1}^n \Filter_{\{x,y,k_i\}}(F)$. This is equal to $\Filter_S(F)$
since $F$ is fixed on $\{x,y\}=\bigcap_{i=1}^n\{x,y,k_i\,\}$.
Finally, $\Filter_{\{x,y,k_i\}}(F)$ is computed as in Example~\ref{ex:filter-on-nuclear-family},
for each $1\le i \le n$.

The algorithm is iterated on all nuclear families until no new genotypes are removed.
If $H'$ is the psmap obtained at the end of the algorithm and $g\in H(i)$ for any $i\in S$,
then $g$ is consistent on all nuclear families to which $i$ belongs.
Let us call $\LG\colon \Maps\to\Maps$
the function that maps an input psmap $H$ to the output psmap $\LG(H)$ according to the
Lange-Goradia algorithm.
In general, $\Filter_I(H) \sqsubseteq \LG(H)$ and the relation may be strict, i.e.,
the algorithm may not eliminate all inconsistent genotypes. 
As shown by Lange and Goradia \cite{LangeGoradia1987}, a sufficient
condition for $\Filter_I(H) = \LG(H)$ is the absence of loops in the
pedigree. As an example, consider the pedigree of Figure~\ref{fig:example2}. The
pedigree contains loops, since there are individuals that mate that have an
ancestor in common (for instance individuals 12 and 13 are both descendant of
individual 8). Therefore, it is not guaranteed that the result of Lange-Goradia
(Figure~\ref{fig:subfigex2lg}) contains only consistent genotypes. In fact,
consider individual 9. Although the genotype $(B,B)$ is not consistent, the
Lange-Goradia algorithm cannot eliminate it. To see that it is not possible to
find a Mendelian fsmap that is $\sqsubseteq$ of that depicted in
Figure~\ref{fig:subfigex2}, consider individual 15. One of his alleles is
$A$. Since his alleles must come from individuals 7, 8, and 9, at least one of
those individuals must have allele $A$. Individuals 7 and 8 do not contain it,
thus 9 must have $A$ as allele, and we can eliminate $(B,B)$. We will see in the
next subsection that the O'Connell and Weeks algorithm is able to eliminate
$(B,B)$ from individual 9.

\subsection{The O'Connell and Weeks algorithm}
\label{sec:ow}

The O'Connell and Weeks algorithm \cite{OConnellWeeks1999} is able to remove all
inconsistent genotypes from a psmap.
The algorithm has the same input of the Lange-Goradia algorithm:
a pedigree $P=\Ped I f m$ and a psmap $H\in\Maps$.
Let us call $\OCW\colon \Maps\to\Maps$
the function that maps an input psmap $H$ to the output psmap $\OCW(H)$ according to the
O'Connell and Weeks algorithm.

First, a suitable set $B\subseteq I$
of \emph{loop breakers} is found. 
A loop breaker is
an individual that is involved in a loop in the pedigree and set
$B$ must contain such an individual for each loop in the pedigree.

A new pedigree $\Clone P = \Ped {I\cup \Clone B} {\Clone f} {\Clone m}$ is built, where
$\Clone B$ contains a new individual $\Clone b$ for each $b\in B$,
$\Clone f$ is undefined for all $\Clone b\in B$, is equal to $f$ for all $x$ such that
$f(x) \not\in B$, and $\Clone f(x) = \Clone {f(x)}$
for all $f(x)\in B$ (and similarly for $\Clone m$).
Thus, $\Clone P$ is obtained from $P$ by breaking all loops.
Then, for each $F\in\Split_B(H)$ a psmap $\Clone F$ on $\Clone P$ is built, where
$\Clone F(x) = F(x)$ for all $x\in I$ and $\Clone F(\Clone b) = F(b)$ for all $b\in B$.
Finally, $\LG(\Clone F)$ is computed for all $\Clone F$
and all output psmaps thus obtained are joined.
Since $\Clone P$ contains no loops,
we have $\Clone F'=\LG(\Clone F) = \Filter_{I\cup \Clone B}(\Clone F)$ for all $\Clone F$.
It is easy to see that it is $\Clone F'(b) = \Clone F'(\Clone b)$ for all $b\in B$ and
that this implies that the restriction of $\Clone F'$ to $I$ is consistent on $I$.
Indeed, if $F'$ is the restriction of $\Clone F'$ to $I$ we have $F'=\Filter_I(F)$.

We note that there is no need to actually build pedigree $\Clone P$,
since $\LG(F)$ will produce the same result as $\LG(\Clone F)$ whenever
$F$ is fixed on $B$. Thus we can simply define
\begin{equation}
  \label{eq:OCW}
  \OCW(H) = \bigsqcup_{F\in\Split_B(H)}\LG(F).
\end{equation}
For each $F\in\Split_B(H)$ we have $\LG(F) = \Filter_I(F)$, thus
we obtain $\OCW(H) = \Filter_I(H)$ from \eqref{eq:split-filter-sup}.

\begin{figure}
  \centering
  \includegraphics[width=0.5\linewidth]{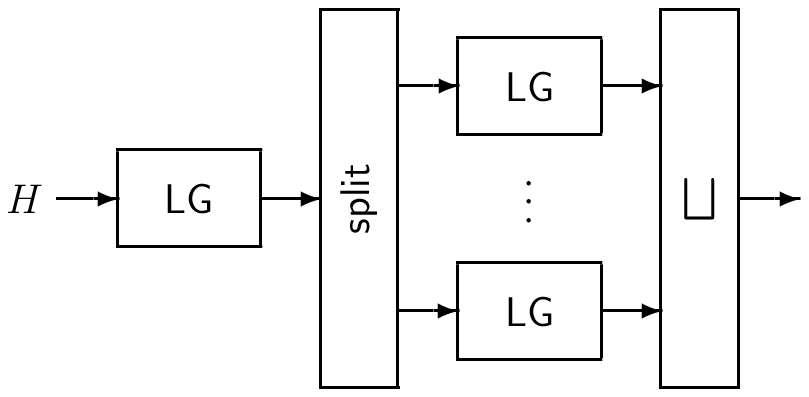}
  \caption{The O'Connell and Weeks algorithm.}
  \label{fig:OCW}
\end{figure}
Fig.~\ref{fig:OCW} shows a block-diagram representation of the O'Connell and Weeks algorithm.
Note that eq.\eqref{eq:OCW} corresponds to the part of the diagram from the $\Split$ block
onwards. The initial $\LG$ block is not necessary for the completeness of the algorithm,
but is introduced in order to try to reduce the cost of the rest of the algorithm,
since the number of Lange-Goradia invocations depends combinatorially on the number
of genotypes assigned to each loop breaker.

\begin{figure}
  \centering
  \subfigure[]{\includegraphics[width=.32\linewidth]{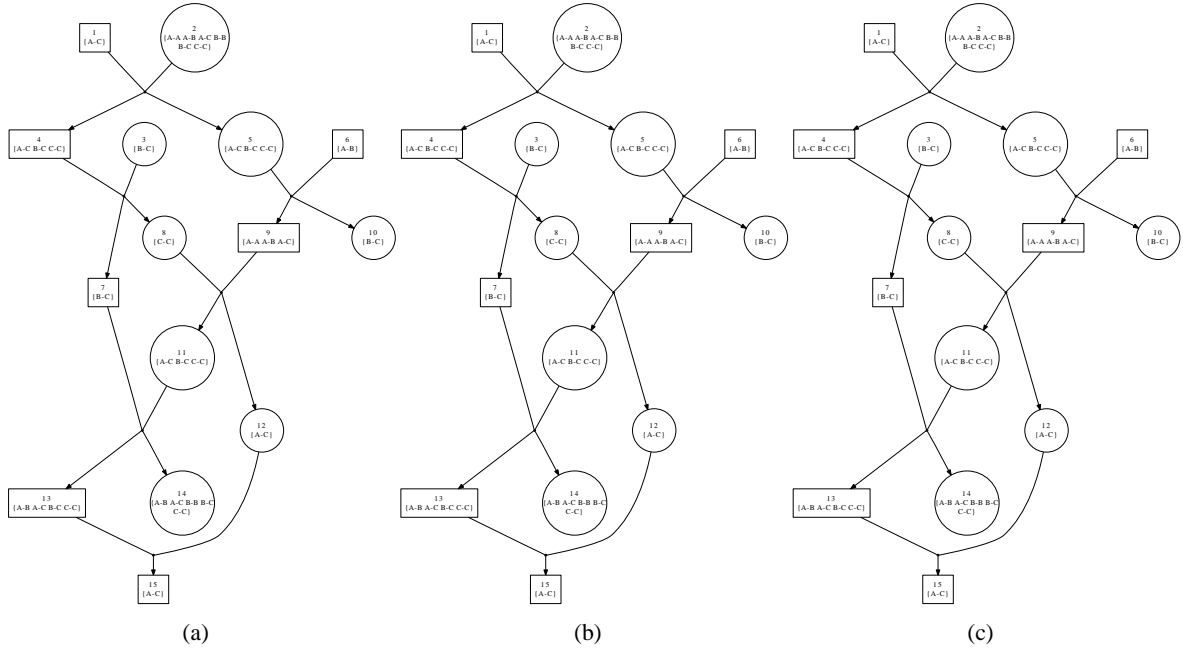}\label{fig:subfigow1}}
  \subfigure[]{\includegraphics[width=.32\linewidth]{example2-ow-1}\label{fig:subfigow2}}
  \subfigure[]{\includegraphics[width=.32\linewidth]{example2-ow-1}\label{fig:subfigow3}}
  \caption{An example of the applications of the O'Connell and Weeks: from the
    pedigree in Fig.\ref{fig:subfigex2lg} the individuals 8 and 12 are chosen as
    loop breakers, leading to three applications of the Lange-Goradia algorithms
    whose results are depicted in this figure.}
  \label{fig:example2-decomposed}
\end{figure}

As an example, consider the pedigree depicted in
Figure~\ref{fig:subfigex2lg}. The pedigree contains various loops that can be
broken, for instance, by choosing individuals 8 and 12 as loop breakers. This
choice leads to three applications of the Lange-Goradia algorithm to the
pedigree of Fig.~\ref{fig:subfigex2lg} in which the individual 12 is typed as
$(A,C)$, $(B,C)$, and $(C,C)$, respectively. The three runs have as results the
psmaps depicted in Figure~\ref{fig:example2-decomposed}. The union of these
three psmaps gives as results the psmap depicted in
Figure\ref{fig:subfigex2ow}. We can note that genotypes $(B,B)$ and $(B,C)$ have
been eliminated from individual 9. 

\section{The {\TheTool} tool}
\label{sec:imple}

We have implemented the O'Connell and Weeks algorithm in  a tool named
\TheTool. {\TheTool} has been developed in C++ and is able to perform genotype
elimination. Using a command-line switch, it is possible to select either Lange-Goradia
or O'Connell and Weeks's algorithm. {\TheTool} receives as input a pedigree in
pre-LINKAGE format, and writes the processed pedigree in a human-readable
form. Moreover, it is also possible to have a DOT-file as output, that can be
processed with Graphviz~\cite{graphviz} to obtain a graphical representation of
the resulting pedigree.

\subsection{Parental information}
\label{sec:pedigr-data-struct}
In the design of our application, we kept the genotypic information separated 
from the parental information. During the parsing of the file, 
parental relations are stored in a redundant set of data structures (list of nuclear families in the
pedigree, list of partners of each individual, list of families each individual
belongs to, etc.). These data structures allow to recover all the parental
relations needed by the consistency algorithms in a fast way. For instance,
during the Lange-Goradia algorithm, to avoid unnecessary iterations, we set up a
working list of the families to be processed. When the genotypes set of an individual
changes, we insert in the working list only the families the individual belongs
to. 

\subsection{Genotypes set as bitmaps}
\label{sec:genotypes-set-as}

Our efficient implementation uses bitmaps to represent elements of
$\powerset(\Genotypes)$ (individual of a psmap). When the set contains few
genotypes, a bitmap needs more space than other alternatives such as binary
search trees. On the contrary, this slight drawback is counter-balanced by many
advantages. First of all, the operations of search, insertion and deletion from
subset of $\Genotypes$ can be completed in constant time. Moreover, when the
maximum number of alleles is known in advance, bitmaps can avoid the use of
dynamic memory, thus speeding up the operations of copy and
allocation/deallocation.  Union and intersection of set of genotypes can be
implemented with bitwise logical operations.  Even the iteration of all the
genotypes in a set can be implemented efficiently by calculating the least
significant bit in a word. 

We chose to represent alleles with unsigned integers in the range $[0,N-1]$,
where $N$ is the maximum number of alleles. With this choice, elements of
$\powerset(\Genotypes)$ are triangular bitmaps with $N$ rows.  When $N=32$, the
$n$-th word of the matrix represents the subset of $\Genotypes$ composed by
genotypes with $n$ as the first allele, and $k<=n$ as the second allele. In this
way, it is easy to build bit masks for manipulating sets of genotypes. 

As an example, consider the optimization suggested in~\cite{OConnellWeeks1999}. To
speed up the initial application of the Lange-Goradia algorithm, O'Connell and
Weeks suggest to pre-process the pedigree by removing those genotypes that can
be easily identified as superfluous by looking at a single parent-child pair. For
instance, when a child is fully specified with alleles $(A,B)$, it is possible
to remove from its parents all the genotypes that do not contain at least one
from $A$ and $B$. With the genotype set represented as a bitmap, it is
sufficient to clear all the bits that are not in words $A,B$ and in columns
$A,B$. The \verb|C++| code of this operation can be found in Figure~\ref{fig:remove}.

\begin{figure}
  \centering
\begin{lstlisting}
void bitmap::reduce_parent_child(int A, int B) {
  // A is always less or equal than B
  unsigned int allele_mask;
  if (A == B) { // homozygous individual
    allele_mask = (1 << A);
    unsigned int i;
    for (i=0; i<A;++i) {
      data[i] &= allele_mask;
    }
    i++; // leave A-th word untouched
    for (; i<32;++i) {
      data[i] &= allele_mask;
    }
  } else {
    allele_mask = (1 << A) | (1 << B); 
    unsigned int i;
    for (i=0; i<A;++i) {
      data[i] &= allele_mask;
    }
    ++i; // leave A-th word untouched
    for (; i<B;++i) {
      data[i] &= allele_mask;
    }
    ++i; // leave B-th word untouched
    for (; i<32;++i) {
      data[i] &= allele_mask;
    }
  }  
}
\end{lstlisting}  
  \caption{The C++ code for the optimization suggested by O'Connell and
    Weeks. When an individual is typed we remove from his/her children (and parents) the
    genotypes that do not contain at least one of his/her alleles. In the code,
    A and B are the alleles of the typed individual.}
  \label{fig:remove}
\end{figure}
  
Concluding, the bitmap has been a key choice for speeding up all the consistency
algorithms.

\subsection{Loop breakers selection}
\label{sec:loop-break-select}
We have seen that the O'Connell and Weeks algorithm executes the Lange-Goradia
algorithm once for every combination of the genotypes of the loop
breakers. Therefore, the selection of loop breakers greatly affects the total
running time of the O'Connell and Weeks's algorithm. In {$\TheTool$}, we chose
to apply the selection strategy suggested by Becker et
al.~\cite{loopBecker}. The idea of the selection algorithm is to prefer to
choose the individuals that break more loops at a time, and to avoid the ones
that have a long list of genotypes. Becker et al. show that this problem is
equivalent to the calculus of the minimum spanning tree of a directed graph. The
graph to be analyzed can be obtained from the parental graph by removing all the
individuals (and corresponding marriage nodes) that do not belong to any
loop. This reduction of the graph must be put in place whenever a new loop
breaker is chosen.  The individuals in this graph are labelled with the result
of a function $f$ that estimates the cost of the selection of the corresponding
loop breaker. The function $f\colon \Maps \times I \to \relset^{+}$ is defined
as $f(H,i) = \log (\sharp H(i)) / d(i)$, where $\sharp H$ denotes the
cardinality of set $H$ and $d(i)$ is the number of neighbours of individual $i$
in the graph. The intended meaning of the function $d$ is to be a heuristic
estimate of the number of loop the individual belongs to. We implemented the
spanning tree calculus with a modified version of the classical algorithm by
Kruskal~\cite{kruskal}. In fact, in this case, the function $f$ (and in
particular $d$) must be recalculated because the graph is reduced whenever a new
loop breaker is found. However, since the cost of selection is only increasing,
the greedy methodology of the spanning tree algorithm can be preserved.

It is easy to see that, by definition of $\Split$, given $S,T \in I$ and $H \in
\Maps$, with $T \subseteq S$ and $H$ fixed on $T$, it holds $\Split_S (H) =
\Split_{S\backslash T}(H)$. Therefore, in the $\Split$ phase, we discard all
the loop breakers that have a single genotype. 

\subsection{Recursive vs non recursive reduction}
\label{sec:recursive-vs-non}
To reduce the number of Lange-Goradia reductions (one for every combination of
the genotypes of the loop breakers), O'Connell and Weeks suggest to use a
recursive version of their algorithm. Instead of calculating all the
combinations and applying the Lange-Goradia reduction, they adopt a backtracking
methodology and execute a Lange-Goradia reduction whenever a loop breaker
genotype is fixed. The algorithm can be
expressed by the following pseudo-code. In the pseudo-code, given a function $f$, we denote with $f\modif{x}{y}$ the
function $f'$ defined as $f'(z)=f(z)$ if $z \neq x$, and $y$ otherwise. This notation is used for updating the  
\begin{algorithm}
  \caption{The recursive version of the O'Connell and Weeks algorithm}
  \begin{algorithmic}[1]
    \STATE $\OCWR$($P$, $B$, $H$) 
    \IF{$B=\varnothing$} 
    \RETURN $H$ 
    \ELSE 
       \STATE $R \leftarrow \bot$ 
       \STATE select an individual $i \in B$ 
       \FOR{$g \in H(i)$}
           \STATE $H' \leftarrow H \modif{i}{g}$ 
           \STATE $R \leftarrow R \sqcup {\OCWR(P, B\setminus i , \LG(H'))}$
       \ENDFOR
       \RETURN $R$
    \ENDIF
  \end{algorithmic}
\end{algorithm}
The rationale behind this approach is to avoid a brute-force exploration of the
results of the $\Split$ function in \eqref{eq:OCW}. However, our experiments
show that this approach does not pay off when coping with large pedigrees and
few combinations to explore. In fact, all the psmaps that are on the recursion
call stack must be initialized and copied, thus leading to an increased use of
memory. When the number of individuals is not high and there are many
combinations to explore, the recursive version is better than the non recursive
one.

\section{Performances of {$\TheTool$}}
\label{sec:performances-thetool}

We have tested {\TheTool} with three different pedigrees. Following the
methodology described in \cite{Pirinen2006}, we have simulated genetic data by
picking founder alleles from the uniform distribution, applying randomly
the Mendelian laws down the pedigree to calculate non-founder alleles, and,
finally, deleting the genotype information of some individuals. 

The first pedigree we considered is composed by 221 individuals. It is a human
pedigree that traces the ancestors of two individuals affected by
hypophosphatasia (HOPS). The pedigree comes from the Hutterite population living
in North America, and it has been used previously in \cite{luolin,Pirinen2006}.

We analyzed 100 datasets for each combination of the number of alleles (5, 7, 10,
12, 15, 17, 20, 25, 30), and of the ratio of untyped individual (5, 10, 20, 30, and 50
percent), for a total of 4500 datasets. 

Then, we tested a larger pedigree composed by 4921 individuals. This pedigree
was also studied in \cite{Pirinen2006} and has been simulated with the method of
Gasbarra et al.~\cite{Gasbarra200575}. It has been used as a benchmark for the
tool Allelic Path Explorer (APE). The pedigree contains 159 founders, and
75 percent of individuals were inbred. Again, simulating genetic data, we have
created 100 datasets for each combination of number of alleles and each
ratio of untyped individuals.

The last pedigree we tested is even bigger. It is composed by 8420 individuals
and has been generated with the tool QMSIM~\cite{qmsim}. It is composed of 10
generations. The founders are 420 individuals (400 females and 20 males).  We
have tested the performance of ${\TheTool}$ on a Intel Core 2 Duo 3.00 GHz
machine equipped with 2GB of RAM and running Ubuntu Linux 9.10 (kernel version
2.6.31-21).

\begin{table}
  \centering
  \begin{tabular}{lrrrr}
    \textbf{Name}&\textbf{Individuals}&\textbf{Generations}&\textbf{\%Founders}&\textbf{Avg
    Family size}\\\hline
    HOPS  & 221  & 12 & 21.72\% & 1.52 \\
    APE   & 4921 & 15 & 3.23\%  & 1.82 \\ 
    QMSIM & 8420 & 10 & 4.99\%  & 2.00 \\\hline
  \end{tabular}
  \caption{The three benchmarks used }
  \label{tab:benchmarkdata}
\end{table}

\begin{figure}
  \centering
  \subfigure[]{\includegraphics[width=.45\linewidth]{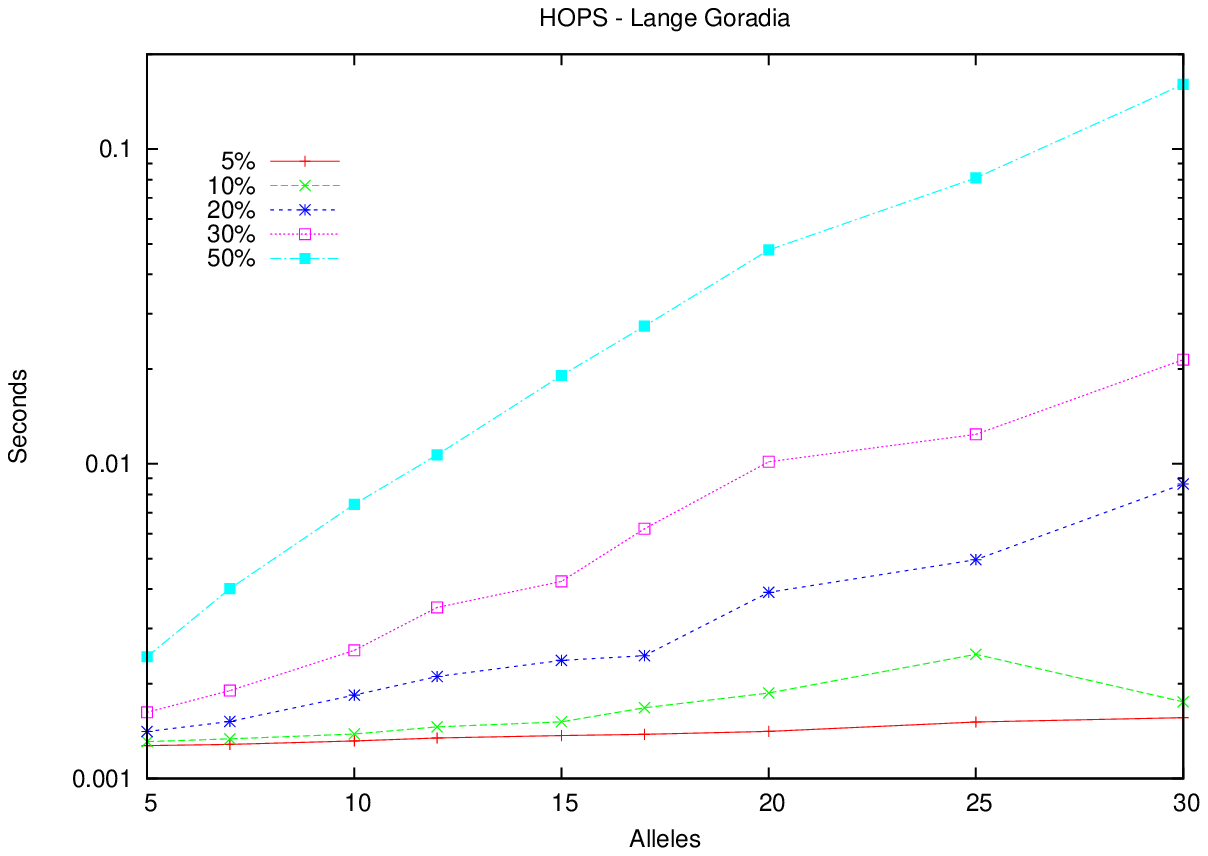}\label{fig:subfighopslg}}
  \subfigure[]{\includegraphics[width=.45\linewidth]{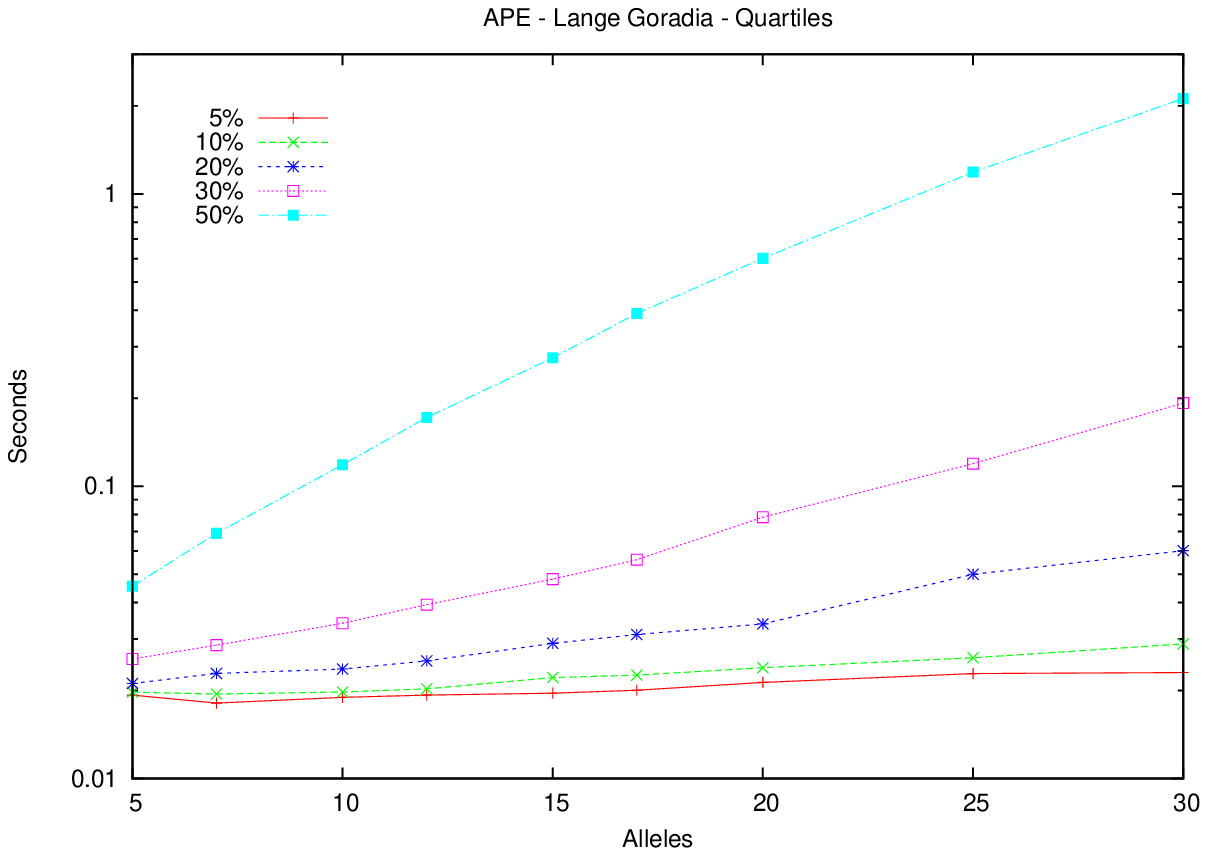}\label{fig:subfigapelg}}\\
  \subfigure[]{\includegraphics[width=.45\linewidth]{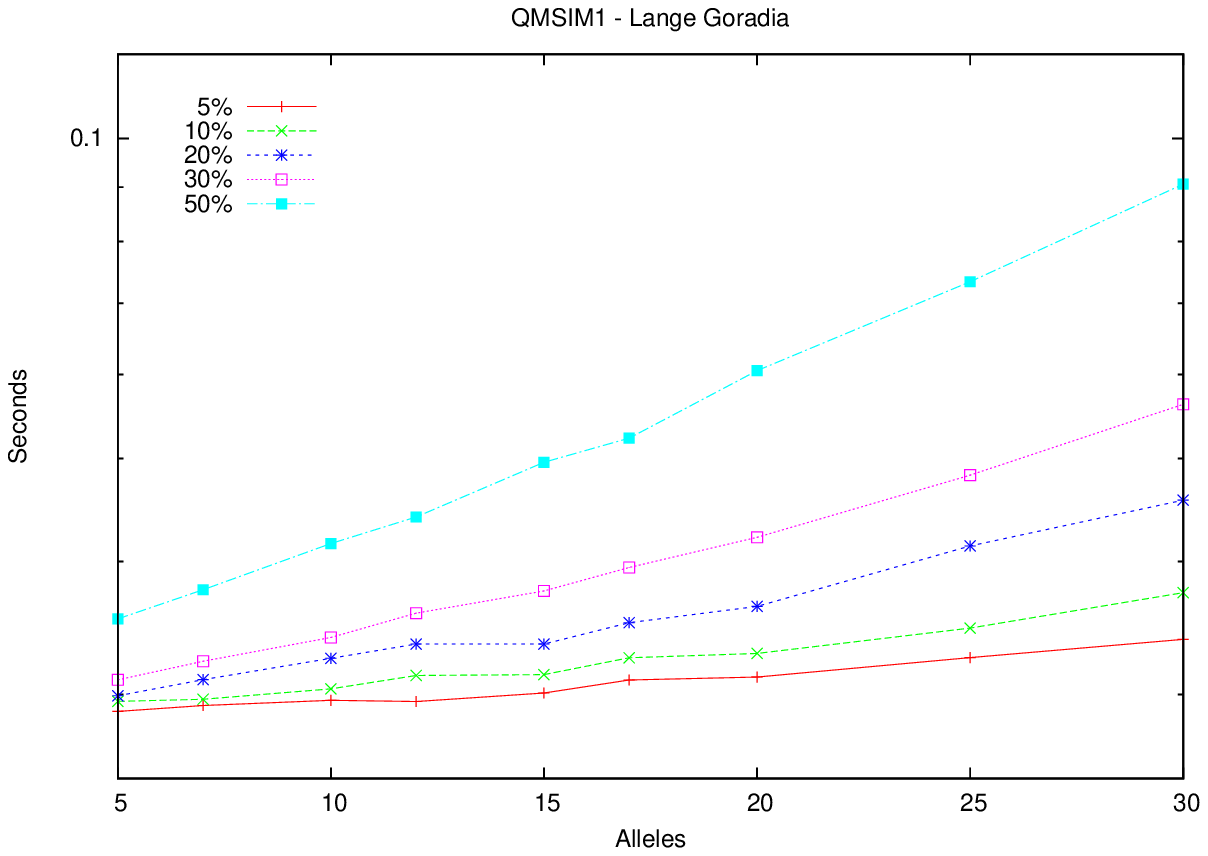}\label{fig:subfigqmsimlg}}
  \subfigure[]{\includegraphics[width=.45\linewidth]{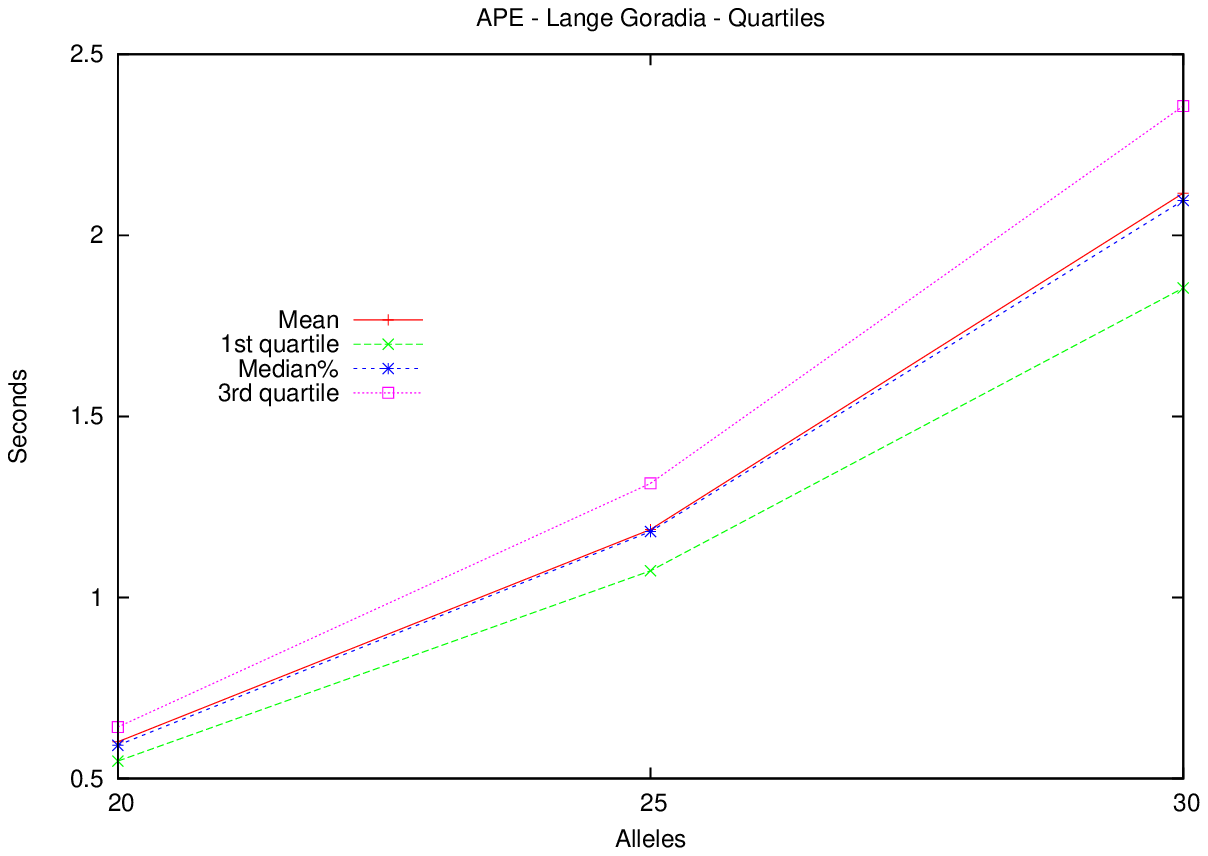}\label{fig:subfigapequartile}}
  \caption{The performance of {$\TheTool$} when the Lange-Goradia algorithm is
    applied: HOPS \ref{fig:subfighopslg}, APE \ref{fig:subfigapelg}, and QMSIM
    \ref{fig:subfigqmsimlg}. In \ref{fig:subfigapequartile} we show the quartile
  for the benchmark APE when the ratio of untyped individuals is set to 50\%,
  where we noticed a significant variance.}
  \label{fig:celer-lg}
\end{figure}

Figure~\ref{fig:celer-lg} shows the execution time of {$\TheTool$} when the
Lange-Goradia algorithm is executed. We have put the number of alleles on the x
axis and there is a line for every percentage of untyped individuals in the
pedigree. Every dot in the graph refers to the average execution time of the 100
datasets for each combination number of alleles-ratio of untyped individual. We
have used a logarithmic scale on the y axis, and therefore the linear trend
corresponds to an exponential growth of the execution time when the number of
alleles is raised. We can note that, even though the QMSIM pedigree is composed
by a larger number of individuals than APE, the execution times are
significantly lower. This could be due to its simple and regular parental
structure (see Table~\ref{tab:benchmarkdata} for a comparison). We have measured
a very low variance among the same 100 datasets, except when the number of
alleles is high and the percentage of untyped individual is set to 50\%. This
effect is particularly evident in benchmark APE. We reported in Figure
\ref{fig:subfigapequartile} the mean, and the first three quartiles of the
execution times of {$\TheTool$}, when the ratio of untyped individuals is 50\%
and the alleles are between 20 and 30.

We have also tested the same benchmarks when {$\TheTool$} executes the O'Connell and
Weeks algorithm. However, in many cases, the loop breakers selection algorithm
is able to find only loop breakers that have a single genotype. In this case, as
we have seen in Section~\ref{sec:loop-break-select}, the O'Connell and Weeks
algorithm is equivalent to the Lange-Goradia. With a low rate of untyped
individuals, the number of loop breakers (from now on we consider only the loop
breakers with more than one genotype) is different from zero only in some
sporadic cases, and thus the average execution time of the O'Connell and Weeks
is very similar to the Lange-Goradia one (the only difference being the loop
breaker selection procedure). We report in Table~\ref{tab:avglb} the average
number of loop breakers and the number of cases generated by the $\Split$
function. As shown in the table, there is the risk of a combinatorial
explosion. When the ratio of unknown individuals has been set to 50\%, we could
not complete the O'Connell and Weeks analysis within 30 minutes of computations
for 3 out of the 900 pedigrees of the HOPS benchmark and 8 out of 900 of the
QMSIM benchmark, and for all the pedigrees of the APE benchmark. 

\begin{table}
  \centering
  \begin{tabular}{lrrrrr}
    \textbf{Benchmark}&\textbf{\% unknown}&\textbf{Avg LB}&\textbf{Max
      LB}&\textbf{Avg Cases}&\textbf{Max Cases}\\\hline 
    HOPS   & $<$50\% & 0.0200 & 2   & 0.04                & 10                   \\
           & 50\%    & 1.7622 & 10  & 686*                & 4.66 $\cdot 10^6$    \\\hline
    QMSIM & $<$50\% & 0.1175 & 6   & 0.18                & 240                  \\
           & 50\%    & 2.6978 & 19  & 955*                & 2.24 $\cdot 10^8$    \\\hline
    APE    & $<$50\% & 0.1175 & 4   & 0.19                & 32                   \\
           & 50\%    & 8.398  & 172 & 4.02$\cdot 10^{67}$ & 3.61 $\cdot 10^{70}$ \\\hline
  \end{tabular}
  \caption{The number of loop breakers and the number of cases generated by the
    ${\Split}$ functions. The mean marked with (*) have been calculated
    excluding testcases with combinatorial explosion (4 for HOPS, 8 for QMSIM).}
\label{tab:avglb}
\end{table}

\begin{figure}
  \centering
  \subfigure[]{\includegraphics[width=.45\linewidth]{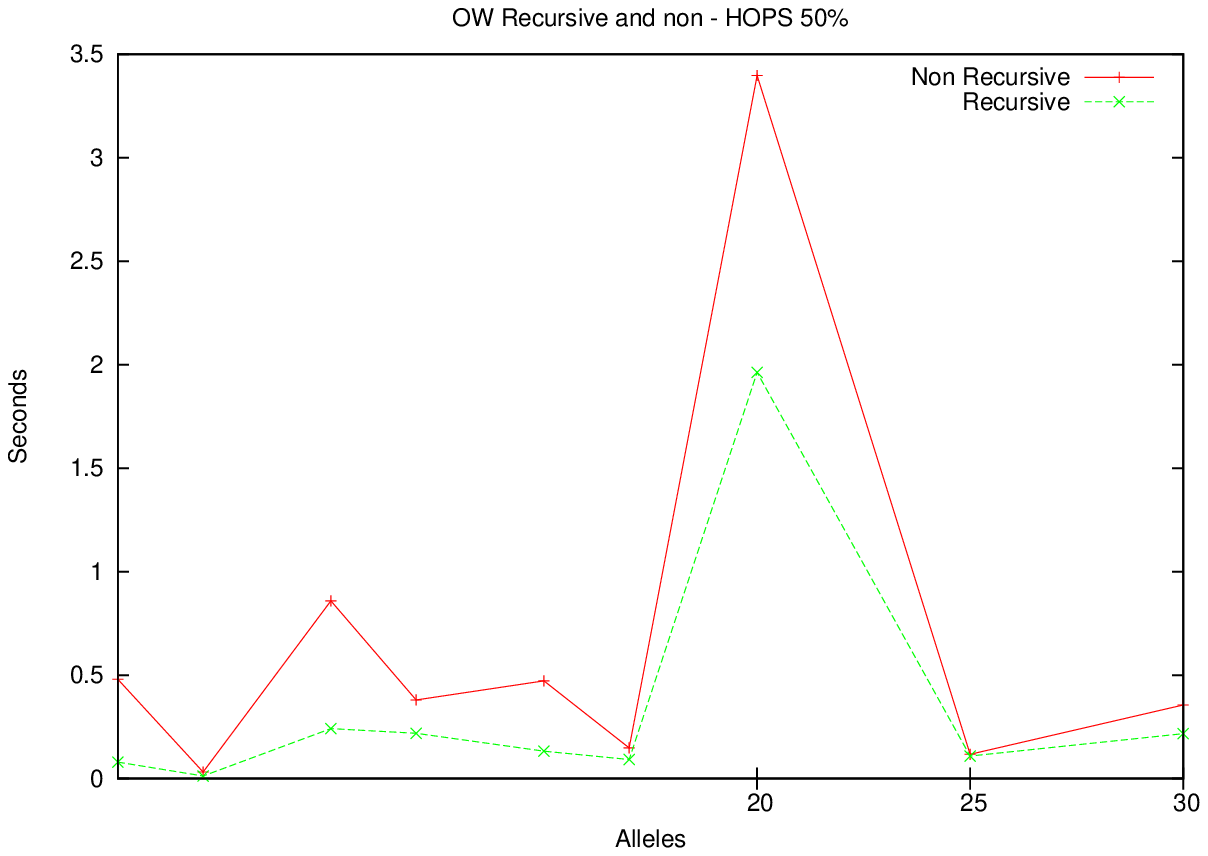}\label{fig:subfighopsow}}
  \subfigure[]{\includegraphics[width=.45\linewidth]{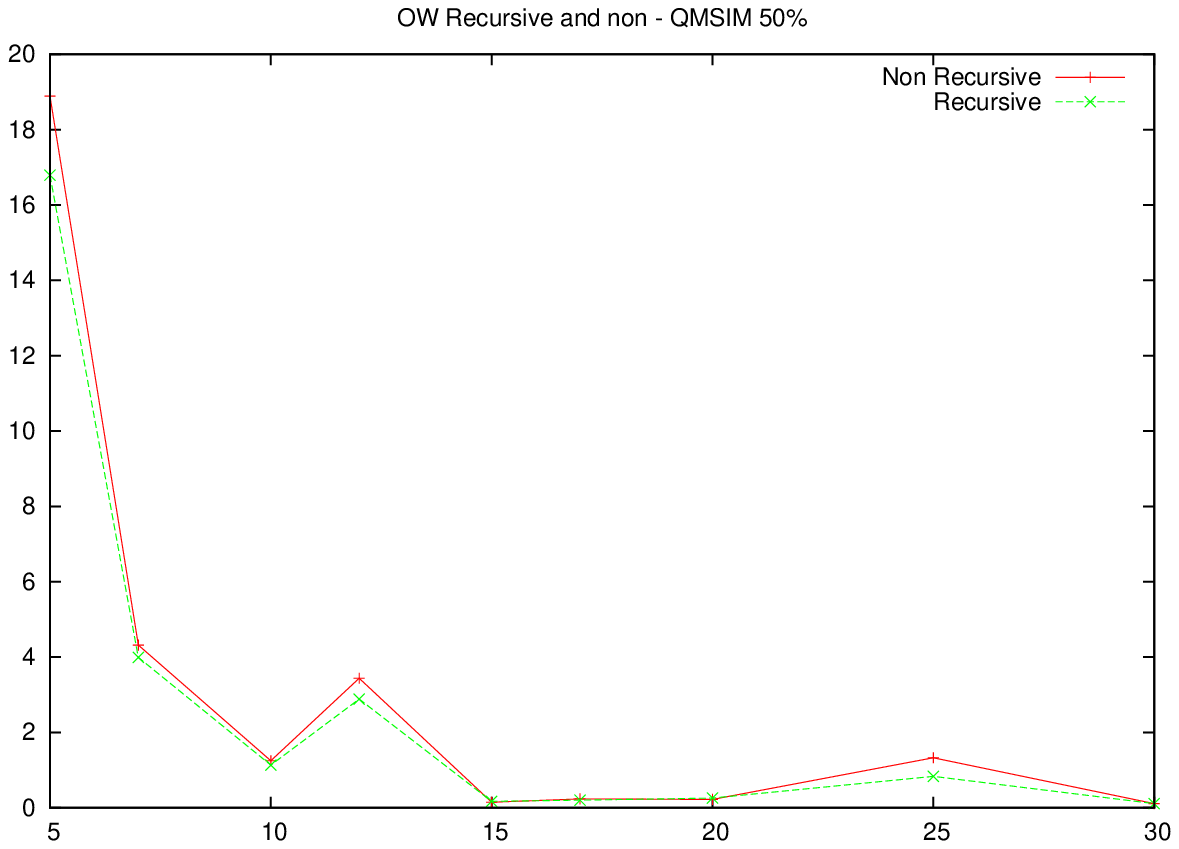}\label{fig:subfigqmsimow}}\\  
  \caption{The execution times of the O'Connell and Weeks algorithm for HOPS
    \ref{fig:subfighopsow}, and QMSIM \ref{fig:subfigqmsimow}, when only half of
  the individuals in the pedigree are typed.}
  \label{fig:owtimes}
\end{figure}

In Figure~\ref{fig:owtimes} we have plotted the average executions of the
O'Connell Weeks algorithm run on the benchmarks HOPS and QMSIM when only half
of the individuals are typed. We can note that the recursive version of the
algorithm dominates the non-recursive version. However, the gap between the twos
is very small in QMSIM, due to the overhead of the backtracking procedure that
nearly counter-balances the advantage of executing fewer Lange-Goradia
iterations. 

\section{Comparison with other software}
\label{sec:comp-with-other}
O'Connell and Weeks have implemented their algorithms in the Pedcheck
program~\cite{pedcheck}. Pedcheck is able to check Mendelian consistency in
pedigree with different levels of accuracy (and therefore with different
computational requirements). Level 1 analysis is able to discover simple errors
related to a single nuclear family (a child and parent's alleles are
incompatible, more than 4 alleles in a sibship, or 3 if there is a homozygous
child). Level 2 correspond to Lange-Goradia algorithm. Level 3 and 4 provide a
basic support to error correction. Level 3 identifies the so-called critical
genotypes (that is the individuals that, if left untyped, make the pedigree
consistent). Level 4 requires to know the frequencies of the alleles to estimate
the most probable corrections. 

\begin{figure}
  \centering
  \subfigure[]{\includegraphics[width=.45\linewidth]{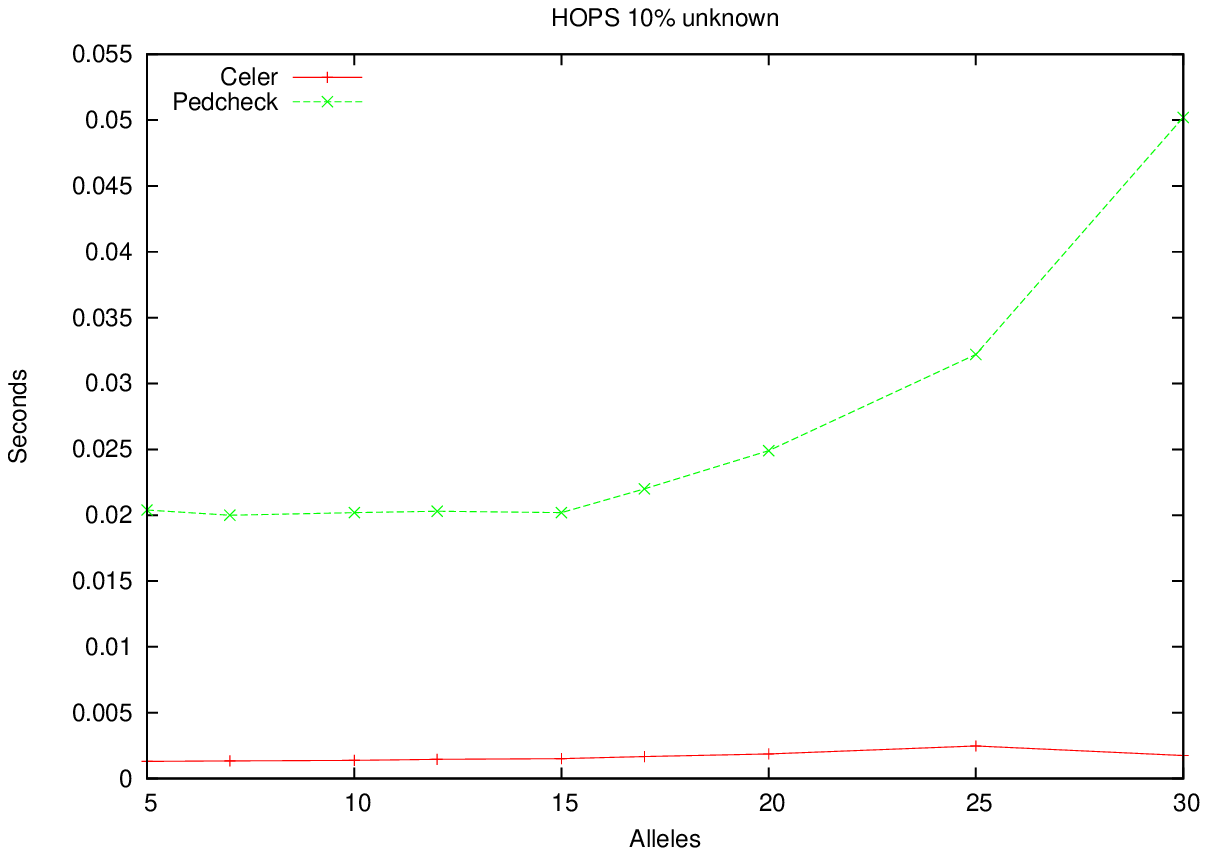}\label{fig:subfigpedcheck10}}
  \subfigure[]{\includegraphics[width=.45\linewidth]{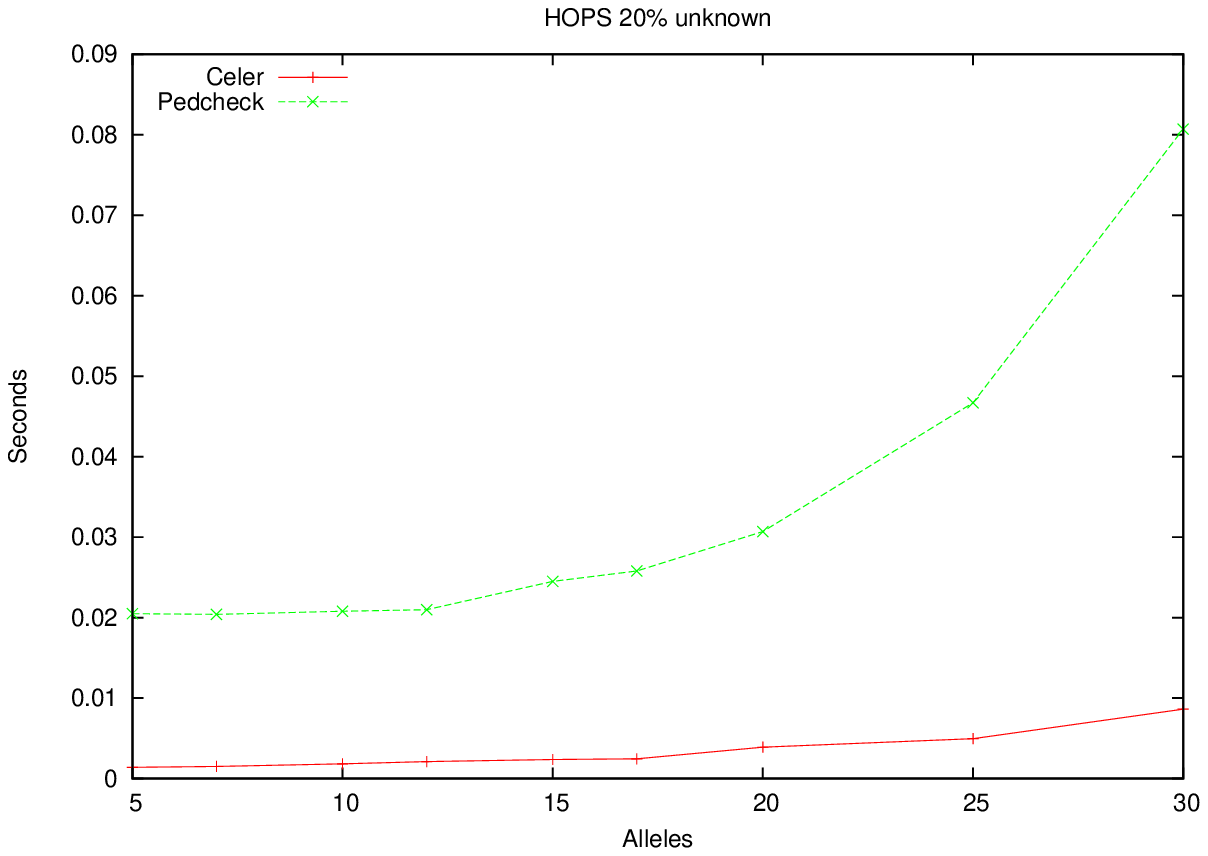}\label{fig:subfigpedcheck20}}\\
  \subfigure[]{\includegraphics[width=.45\linewidth]{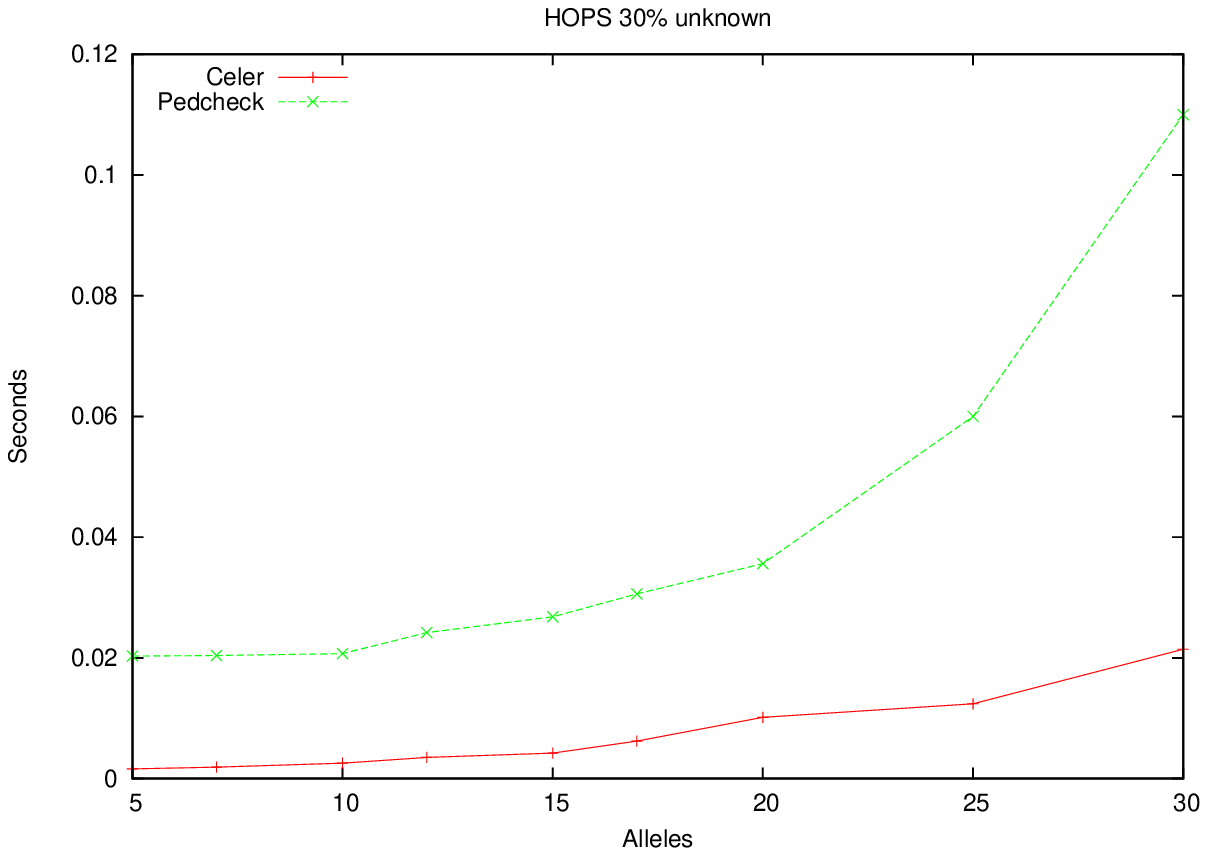}\label{fig:subfigpedcheck30}}
  \subfigure[]{\includegraphics[width=.45\linewidth]{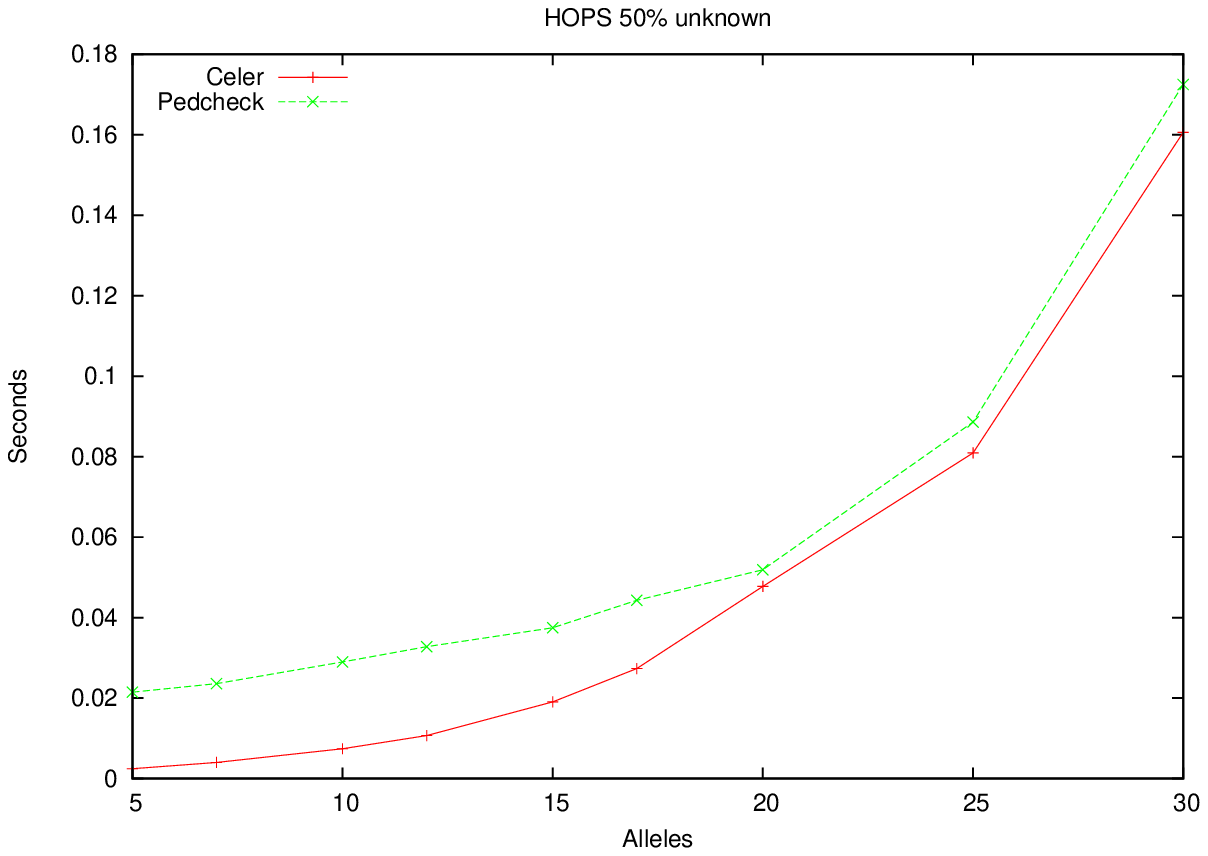}\label{fig:subfigpedcheck50}}
  \caption{Comparison with Pedcheck}
  \label{fig:pedcheckcfr}
\end{figure}

At this time {\TheTool} is more precise than Pedcheck as regards to
genotype elimination, but it does not offer error correction capabilities. 
{\TheTool} is more precise because it can also perform O'Connell and Weeks
algorithm that we have seen is more precise than the Lange-Goradia algorithm.
Moreover, when Pedcheck is applied to large pedigrees, even the Level 2
(Lange-Goradia) phase, takes a considerable amount of time. For example,
consider the QMSIM benchmarks (8420 individuals and 4000 families). Even with
only 10\% of untyped individuals and 5 alleles, Pedcheck needs about 10 minutes
of computation, while our program executes the Lange-Goradia algorithm in less
than 20 milliseconds. We performed the same tests that we used on our tool and
we found that Pedcheck could complete the analysis in times comparable with ours
only on the HOPS benchmark . We report in Figure~\ref{fig:pedcheckcfr} the
average execution times of {$\TheTool$} (with the Lange-Goradia algorithm) and
Pedcheck (level 2 analysis) for the HOPS benchmarks and ratio of untyped
individuals varying from 10 to 50\%. We can see that {$\TheTool$} always
outperforms Pedcheck.

Mendelsoft~\cite{Sanchez2008} is another tool that is able to check Mendelian
consistency and perform error correction. Sanchez et al. model the Mendelian
consistency problem with soft constraint networks and use a generic weighted
constraint network (WCN) solver. In this way, they are not limited to a single error
and can also correct pedigree with multiple errors. They evaluate their tool
with random and real pedigrees composed of thousands of individuals and
containing many errors. Even if we cannot directly compare Mendelsoft with
{$\TheTool$} (that does not have error correction capabilities), we can note
that the memory requirements of the WCN solver are very high. We have tested
Mendelsoft with a machine equipped with 2GB of RAM and in many cases the program
crashed because the amount of virtual memory was not sufficient. In particular,
for the HOPS pedigree, Mendelsoft do not complete with this amount of RAM when
the number of alleles is above 12.

\section{Conclusions and future works}
\label{sec:concl-future-works}

We have described the design and implementation of {$\TheTool$}, a program that
performs genotype elimination. The design of the program has been aided by a
formal description of the problem that highlighted the critical aspects of the
algorithms and helped us to find the best data structures. We have measured the
performances of the program and we have found that {$\TheTool$} is able to cope
with large pedigrees. In the future, we would like to improve the working list
selection algorithm of the Lange-Goradia elimination procedure and to test
different loop breakers selection algorithms on highly-looped pedigrees, such
as the one found in the APE test cases.



\bibliographystyle{eptcs} 
\bibliography{PEDIGREE}

\end{document}